\documentclass[slac_one]{revtex4}
\usepackage{graphicx}
\usepackage{fancyhdr}
\usepackage{amssymb,amsmath,epsfig,subfigure}
\begin{document}

\title{Top Quark Phenomenology}
\author{Rikkert Frederix} \affiliation{Institute for Theoretical
  Physics, University of Zurich, Zurich, Switzerland}

\begin{abstract}
  In this talk three 2-sigma deviations from the Standard Model
  predictions in the top quark sector are briefly discussed. These are
  the excess of events in the tail of the $H_T$ distribution in
  $t\bar{t}$ events, the top-quark charge asymmetry and the
  discrimination of s- and t-channel events in single top. The latter
  has only been observed by CDF, while the other two are found by both
  CDF and D\O.
\end{abstract}

\maketitle

\thispagestyle{fancy}

\section{TOP QUARKS AT THE TEVATRON}
Everything we know about the top quark is coming from the experiments
at the Tevatron collider at Fermilab. Since its discovery in
1995~\cite{Abe:1995hr,Abachi:1995iq}, many of its properties have been
firmly established over the years. The most precise measurements are
its mass~\cite{:1900yx} and cross section~\cite{note9913,note6038},
but results for other observables and properties of the top quark are
improving as well. This includes: the top pair invariant mass
distribution~\cite{Aaltonen:2009iz,note5882}, the forward-backward
charge asymmetry~\cite{note10185,note6062}, the top quark width and
its lifetime~\cite{note10035,note6034}, the branching fraction to $Wb$
compared to $Wq$~\cite{Acosta:2005hr,Abazov:2008yn}, the $W$-boson
helicity fractions~\cite{note10211,note5722}, spin correlation in the
top quark pair production~\cite{note10211,note5950}, the top pair
production mechanism ($gg$ versus $q\bar{q}$ initial
state)~\cite{:2007kq}, the top quark's electric
charge~\cite{note9939,Abazov:2006vd}, the mass difference between top
and anti-top quarks~\cite{note10173,Abazov:2009xq}, the total single
top cross section~\cite{Group:2009qk} including bounds on the CKM
matrix element $|V_{tb}|$~\cite{Group:2009qk} and discrimination
between the single-top s- and t-channel production
mechanisms~\cite{Aaltonen:2010jr,Abazov:2009pa}. Furthermore there
have been many searches for new physics in the top quark sector, such
as: resonant contributions to the $t\bar{t}$
production~\cite{Aaltonen:2009iz,note5882}, search for fourth
generation $t'$ and $b'$ quarks~\cite{note10110,note5892}, scalar top
admixture in $t\bar{t}$ events~\cite{Abazov:2009ps}, top decays to
charged Higgs bosons~\cite{note10104,:2009zh}, $W'$ and $H^+$ resonant
contributions to s-channel single
top~\cite{Aaltonen:2009qu,Abazov:2008vj,Abazov:2008rn}, anomalous
$Wtb$ couplings~\cite{Abazov:2009ky}, production via a FCNC
($u(c)+g\to t$)~\cite{Aaltonen:2008qr,Abazov:2010qk} and decay in
$t\to Zq$~\cite{:2008aaa}.

All measurements are in agreement with the Standard Model expectations
within uncertainties. There are, however, three measurements that are
slightly off: in the search for $t'$ quarks, the tail of the $H_T$
distribution has a small excess of events, the top quark charge
asymmetry is more pronounced than expected and the discrimination of
the s- and t-channel cross section in single top is off by more than 2
sigma (CDF only).  In the rest of this talk I'll give my personal view
on these three experimental results.

\section{THE THREE 2-SIGMA DEVIATIONS}
\subsection{The Tail Of The $H_T$ Distribution}
\begin{figure*}[h]
\centering
    \mbox{
      \subfigure[]{
        \epsfig{file=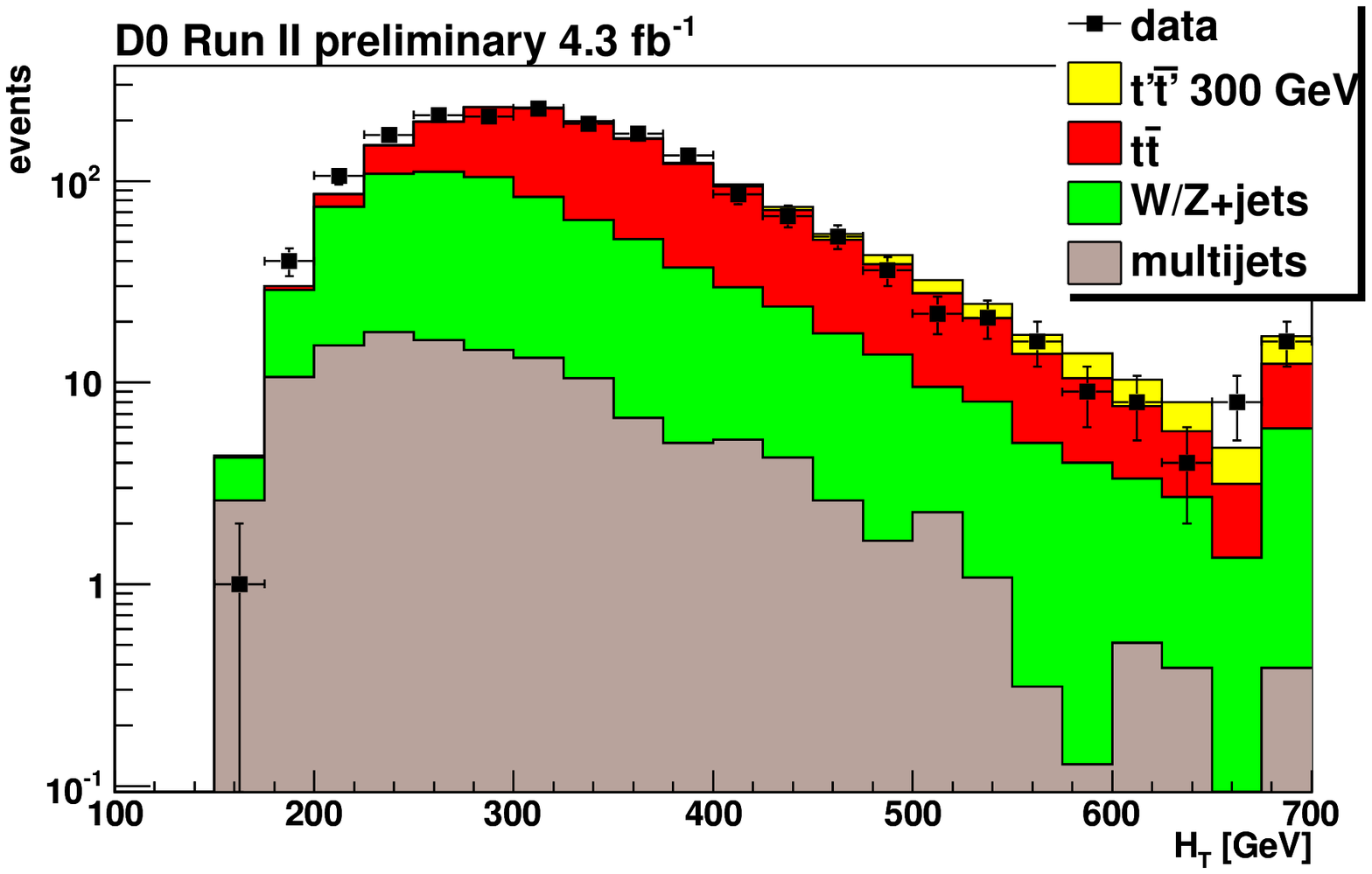, height=5cm}
      }
      \subfigure[]{
        \epsfig{file=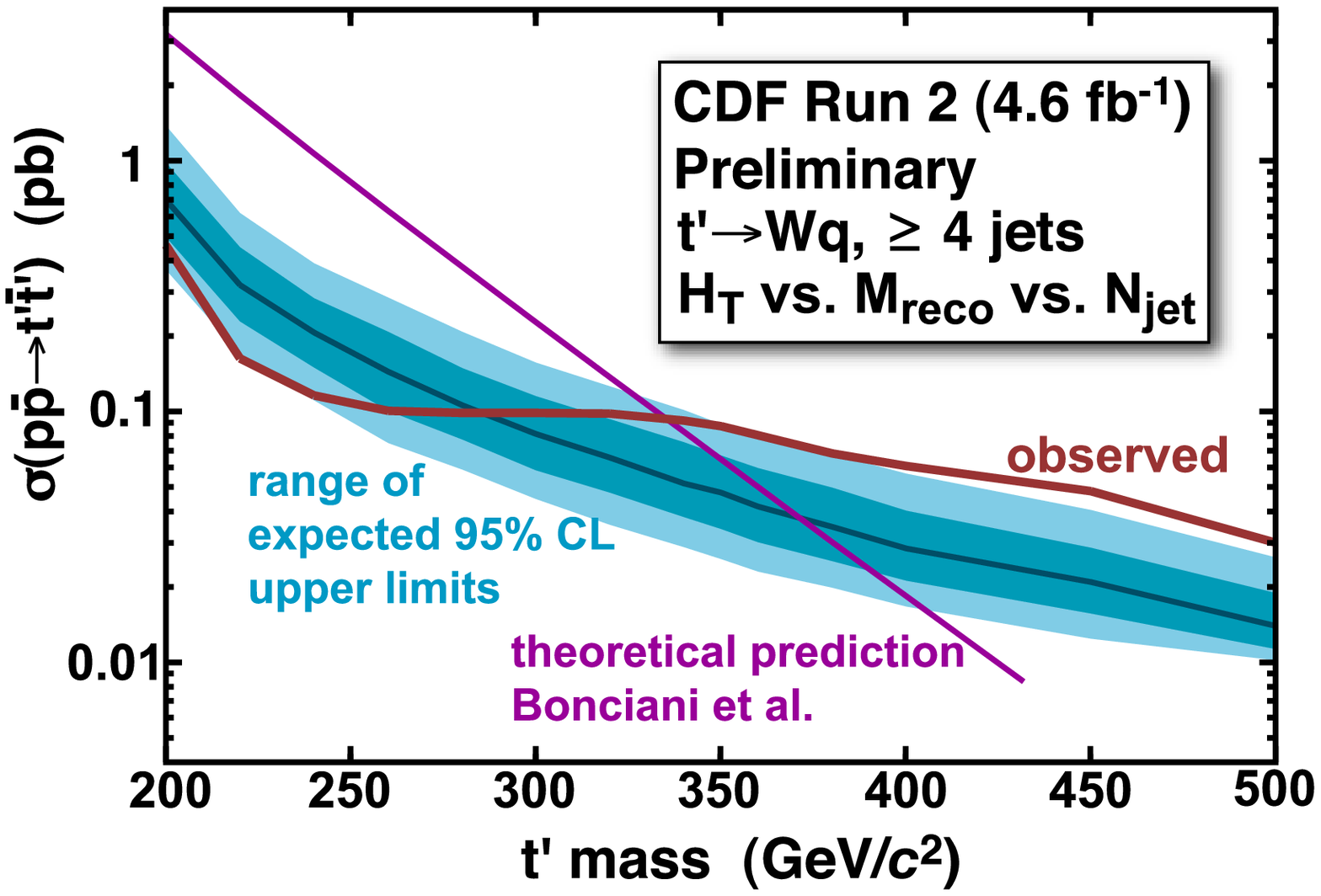, height=5.2cm}
      }
    }
    \caption{The $H_T$ distribution as measured by the
      D\O~collaboration (a).  Observed (red curve) and expected (blue
      band) limits on the $t'$ pair production cross section as a
      function of the $t'$ mass as extracted from the data by CDF are
      shown (b). Figures are taken from
      Ref.~\protect\cite{note5892}~and~\protect\cite{note10110},
      respectively.} \label{fig1}
\end{figure*}

The $H_T$ observable is defined as the scalar sum of all transverse
energies of the jets, leptons and missing $E_T$ and is a measure for
the overall scale of the process. Recently both CDF and D\O~have
updated their analysis for the search of heavy fourth generation
quarks, in particular in $t'$, i.e.~a fourth generation top quark,
that decays predominantly to $W^+b$. After selection cuts, the $H_T$
distribution shows an excess of events in the tail of the
distribution. In fig.~\ref{fig1}(a), this distribution is shown for
D\O. As can be clearly seen from the last two bins (where the final
bin is an overflow bin and contains also all events with
$H_T>700\textrm{ GeV}$) there are slightly more events than could have
been expected from $t\bar{t}$, $W/Z+\textrm{jets}$ and multi-jet
backgrounds. The inclusion of a heavy $t'$ quark to the physics model,
increases the number of events in the tail of the distribution and
results in a better description of the data. This can also be seen
from the plot in fig~\ref{fig1}(b), where for a $t'$ mass larger than
about 350 GeV the observed limit is at the border of the 95\%
confidence level for the predicted limit. This means that limit on the
cross section for $t'$ pairs is much milder than expected, which
reflects the slight excess of events found in the tail of the $H_T$
distribution.

From a theoretical point of view, the tail of the $H_T$ distribution
is difficult to model, because it is an observable quite sensitive to
higher order corrections, see
e.g.~Ref.~\cite{Rubin:2010xp}. Furthermore, the approximation of
describing extra radiation by a logarithm, as is done in all parton
showers, does not capture hard radiation correctly, see
e.g.~Ref.~\cite{Alwall:2008qv}. Using a CKKW~\cite{Catani:2001cc} or
MLM~\cite{Alwall:2007fs} technique to merge higher multiplicity matrix
elements with the parton shower could help in describing the
process. This has been done for the $W+\textrm{jets}$ background, but
not for the $t\bar{t}$ final state and therefore this might have an
impact on the predictions and increase the number $t\bar{t}$ events in
the tail and explain the slight excess.

Given that the effect is small (2 sigma) and non-trivial to model by
Monte Carlo, it is not (yet) significant. More data, and in particular
the LHC with its larger center-of-mass energy, and more refined
analyses using matrix-elements with parton-shower matching will shed
light on these effects.

\subsection{Top Quark Charge Asymmetry}
In pure QCD production of top quark pairs at hadron colliders, the
lowest order prediction shows no preferred direction for the top and
anti-top quarks. When the next-to-leading corrections are taken into
account, the interference between gluon radiation from the initial
state quark line and the top quark line in the $q\bar{q}\to t\bar{t}g$
channel has a negative contribution to the charge asymmetry, while the
contribution from the interference between the virtual box diagrams
and the Born is positive. Furthermore there is a negligible
contribution to the charge asymmetry from the flavor excitation
processes $qg\to t\bar{t}q$. Top quark production by gluon fusion is
symmetric in its initial state and therefore does not contribute to
the asymmetry.

Quantitatively, the contribution from the virtual corrections is larger,
hence the top quarks prefer to go in the direction of the incoming
quark, and the anti-top quarks in the directions of the incoming
anti-quark. Including the (LO) electroweak effects for the Tevatron
the charge asymmetry is given by~\cite{Kuhn:1998jr,Kuhn:1998kw}
\begin{align}\label{ca.1}
A_{FB}(\textrm{lab})&=0.051\pm 0.006,\nonumber\\
A_{FB}(t\bar{t})&=0.078\pm 0.009,
\end{align}
where the uncertainties are coming from renormalization and
factorization scale dependence and
\begin{align}
A_{FB}(\textrm{lab})&=\frac{\int_{y>0}N_t(y)-\int_{y>0}N_{\bar{t}}(y)}
{\int_{y>0}N_t(y)+\int_{y>0}N_{\bar{t}}(y)}\nonumber\\
A_{FB}(t\bar{t})&=\frac{\int N(\Delta y>0)-\int N(\Delta y<0)}
{\int N(\Delta y>0)+\int N(\Delta y<0)},
\end{align}
are the definition of the charge asymmetry in the lab and the
$t\bar{t}$ rest frames, respectively.  Here, $N_{t(\bar{t})}(y)$ is the
number of top (anti-top) quarks as a function of the rapidity $y$ and
$\Delta y=y_t -y_{\bar{t}}$ the difference in rapidities between the
top and anti-top quarks.

Recently, the prediction for the asymmetry in the top pair rest frame
has been improved by including threshold logarithms at all orders and
it was found that the results are quite stable under inclusion of 
these all-order effects~\cite{Almeida:2008ug,Ahrens:2010zv},
\begin{equation}\label{ca.2}
A_{FB}(t\bar{t})=0.073^{+0.009}_{-0.007}\,.
\end{equation}

Both the CDF and D\O~collaborations measure a non-zero top quark
charge asymmetry. CDF finds the following values~\cite{note10185}
\begin{align}
A_{FB}(\textrm{lab})&=0.073\pm 0.028 \quad\textrm{CDF}\nonumber\\
A_{FB}(t\bar{t})&=0.057\pm 0.028 \quad\textrm{CDF},
\end{align}
while D\O~has only performed the analysis for the asymmetry in the top
pair rest frame~\cite{note6062}
\begin{equation}
A_{FB}(t\bar{t})=0.08\pm 0.04\textrm{ (stat.)}\pm 0.01\textrm{ (syst.)}
\quad\textrm{D\O}.
\end{equation}

However, these results are uncorrected for hadronization, underlying
event, background effects, etc.~which makes a direct comparsion with
the predictions, eq.~(\ref{ca.1}) and (\ref{ca.2}), unfair. The CDF
collaboration provides also results corrected for these
effects~\cite{note10185}, which enhances the the asymmetry,
\begin{align}
A_{FB}(\textrm{lab})&=0.150\pm 0.050\textrm{ (stat.)}\pm 0.024\textrm{ (syst.)}
\quad\textrm{CDF (corrected)}\nonumber\\
A_{FB}(t\bar{t})&=0.158\pm 0.072\textrm{ (stat.)}\pm 0.017\textrm{ (syst.)}
\quad\textrm{CDF (corrected)}.
\end{align}
These measurements are almost two standard deviations larger than the
theory predictions given in eq.~(\ref{ca.1}) and (\ref{ca.2}). Notice
that the uncorrected results from CDF and DO~are in agreement with
each other, and if we assume that the unfolding to parton level is the
same for D\O~as it is for CDF, also the asymmetry as measured by
D\O~is also about 2 sigma from its predicted value in perturbative
QCD.

Even though the effect is around two standard deviations, and could
therefore very well be a statistical effect, from the theory point of
view many models beyond the SM were studied to see if these effects
could be explained by new physics, see e.g.~Ref.~\cite{Cao:2010zb} and
references therein. However, before assigning the large measured
asymmetry to BSM effects, another option should be considered as
well. Even though the NLO (i.e.~first non-zero) predictions are stable
under all-order resummation of threshold logarithms, this does not
take into account that there might be a sizable effect from higher
order virtual corrections, i.e.~the two loop corrections to $q\bar{q}
\to t\bar{t}$. The need to go from NLO to the full NNLO is
acknowledged.

The ingredients to go one full order higher in perturbation theory are
almost all known in analytic
form~\cite{Bonciani:2008az,Bonciani:2009nb,Korner:2008bn,Dittmaier:2007wz}
and the remaining contributions can be computed
numerically~\cite{Czakon:2008zk}. The one important missing ingredient
is the prescription to combine the contributions in an infra-red
finite way. Several methods have been
proposed~\cite{GehrmannDeRidder:2005cm,Catani:2007vq,Czakon:2010td,Somogyi:2005xz},
but non of them have been proven to work for top pair production in
hadron collisions so far. Obtaining these results needs further
theoretical developments.

From the experimental point of view the measurements are still
statistically dominated, so with more data the measurements will
become more precise. Due to the fact that the LHC as a proton-proton
collider and that the initial state is dominated by gluons, measuring
the asymmetry at this collider will be channeling at the least.

\subsection{Discrimination Between s- And t-channel Events By CDF}
Single top quark production has been firmly established with 5
standard deviations last year~\cite{Abazov:2009ii,Aaltonen:2009jj}. At
the Tevatron only two of the three single top production channels
contribute significantly, i.e.~the s channel (with a cross section
around 1 pb) and the t channel (about 2 pb). The total single top rate
measured at the Tevatron agrees very well with the sum of the s- and
t-channel cross section predictions.

\begin{figure*}[h]
\centering
    \mbox{
      \subfigure[]{
        \epsfig{file=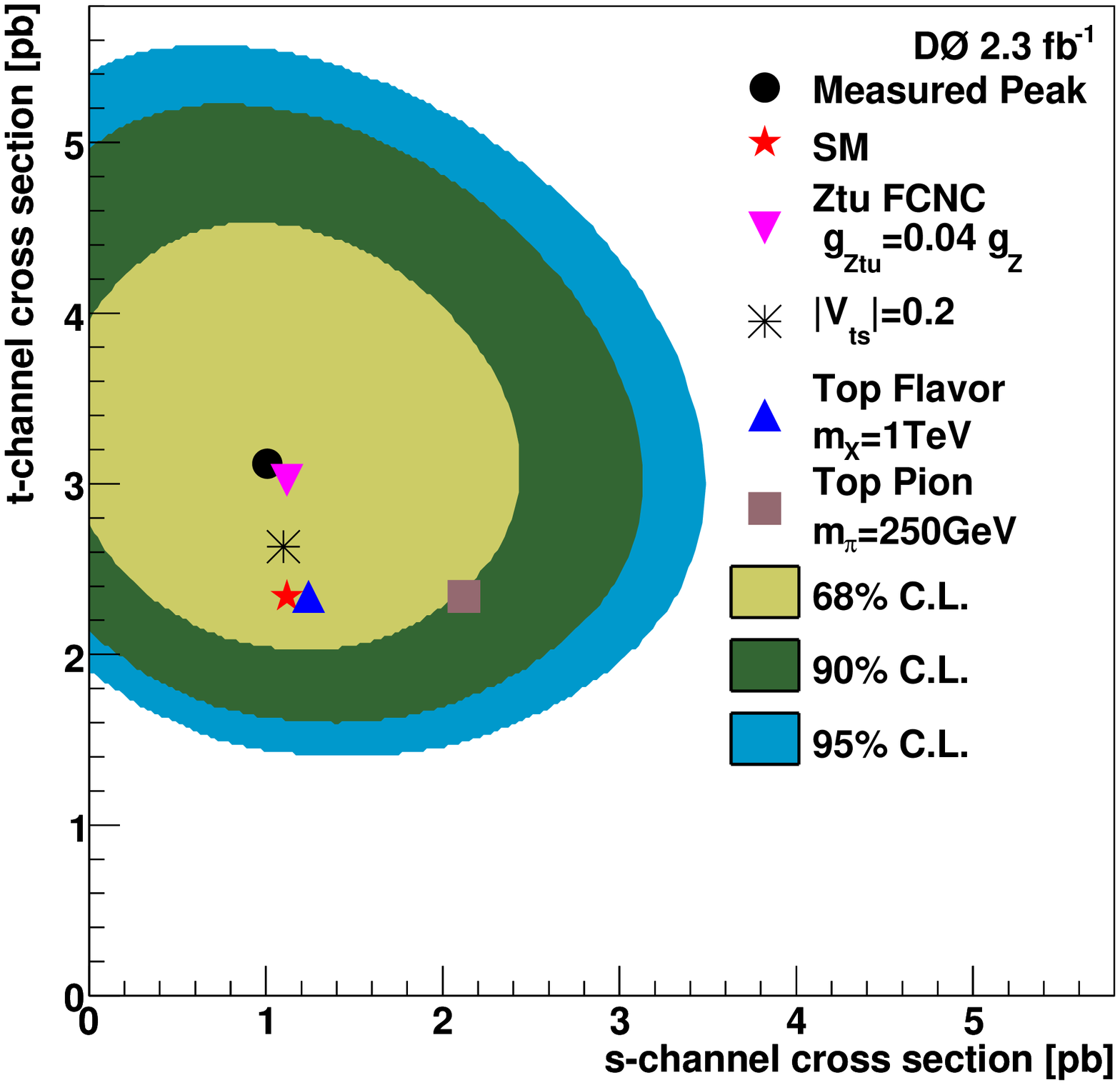, height=5cm}
      }\qquad\qquad\qquad
      \subfigure[]{
        \epsfig{file=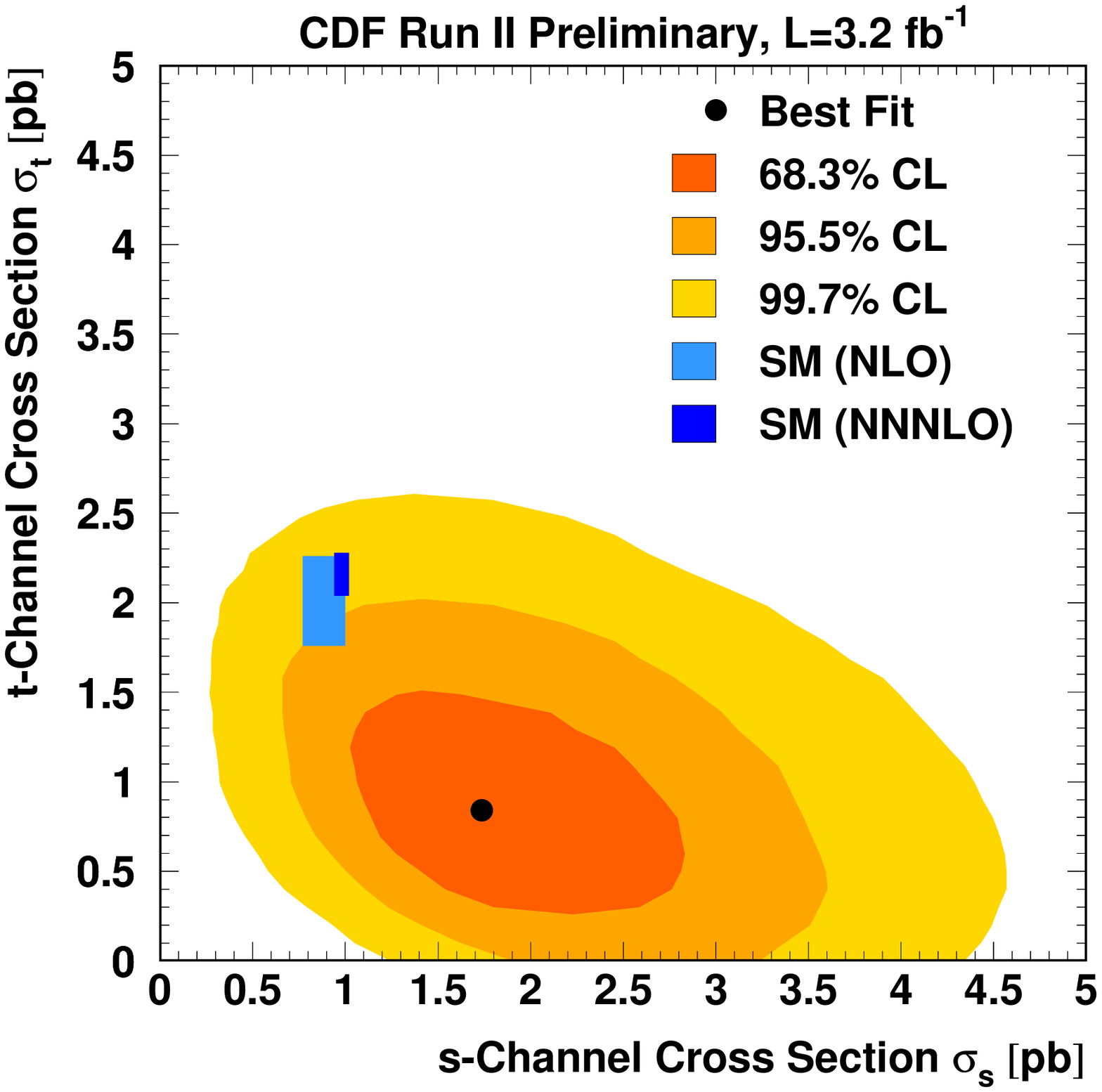, height=6cm}
      }
    }
    \caption{Posterior probability for s- and t-channel single top
      production by the D\O~(a) and CDF (b) collaborations. Figures
      are taken from
      Ref.~\protect\cite{Abazov:2009pa}~and~\protect\cite{Aaltonen:2010jr},
      respectively.} \label{fig2}
\end{figure*}

For D\O~also the discrimination between the s- and t-channel cross
section agrees well within 1 sigma with the theoretical predictions,
see fig.~\ref{fig2}(a). For CDF the situation is worse: the
theoretical predictions are just outside the 95.5\% C.L.~contour, as
can be seen in fig.~\ref{fig2}(b). This discrimination is made using
multi-variate techniques that use as much information as possible from
the event topologies~\cite{Aaltonen:2010jr,Abazov:2009pa}.

\begin{figure*}[b]
\vspace{-10pt}
\centering
    \mbox{
      \subfigure[]{
        \epsfig{file=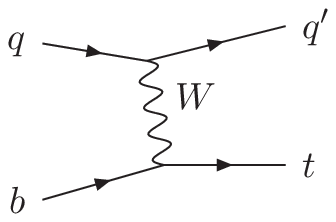, height=2.2cm}
      }\qquad\qquad
      \subfigure[]{
        \epsfig{file=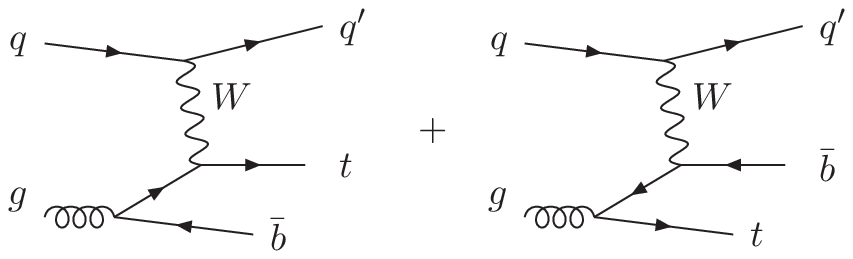, height=2.2cm}
      }
    }
    \caption{Lowest order Feynman diagrams for the t-channel
      single-top process in the five (a) and four (b) flavor
      schemes.} \label{fig3}
\end{figure*}

It is known that the t-channel process, see fig.~\ref{fig3}(a), is
difficult to model by Monte Carlo event generators due to the initial
state $b$ quark. After showering of the events, the initial state b
quark is modeled (in most cases) to come from an gluon splitting into
a $b\bar{b}$ pair, for which the $\bar{b}$ quark becomes a final state
parton, which is called the spectator-$b$ in the following. When only
using the $2\to2$ process as in fig.~\ref{fig3}(a), the transverse
momentum of the spectator-$b$ is greatly underestimated by the parton
shower. This has been acknowledged and addressed by applying a merging
procedure for leading order $2\to2$ and $2\to3$
events~\cite{Boos:2006af}. There is much freedom in the precise choice
of the merging which can lead to a large spread in predictions for the
spectator-$b$.

Recently, the NLO corrections to t-channel single top production in
the 4-flavor scheme, i.e.~starting from the $2\to3$ process,
fig.~\ref{fig3}(b), were
performed~\cite{Campbell:2009gj,Campbell:2009ss}. This calculation
gives the distributions for the spectator-$b$ quark for the first time
at NLO, and should therefore be considered the predictions to validate
the merging procedure of the $2\to2$ and $2\to3$ MC samples against.

Comparing the new predictions for the fraction of spectator-$b$ quarks
with large transverse momentum ($p_T>20\textrm{ GeV}$) and small
pseudo rapidities ($|\eta|<2$) it has been found that results from the
matching procedure as used by the D\O~collaboration follows the NLO
predictions quite well~\cite{Binoth:2010ra}. Unfortunately, for the
Monte Carlo samples used by the CDF collaboration, which are based on
the ZTOP program~\cite{Sullivan:2004ie}, this ratio is about half the
size as the NLO prediction~\cite{lueck}. This means that CDF assumed
that the t-channel events have only about half as many spectator-$b$
quarks as predicted by NLO. Given that s-channel events have an extra
$b$ quark compared to the $2\to2$ t-channel events, at first sight
this discrepancy could explain the measured excess of s-channel events
compared to t-channel events: many t-channel events with a
spectator-$b$ quark in the data were assigned to belong to the
s-channel cross section. However, a careful improvement in the
analysis by the CDF collaboration taking into account the new NLO
computation, shows that this effect on the discrimination of the s-
and t-channel cross sections is negligibly
small~\cite{lueck}. Therefore, even with the improved theoretical
predictions the 2 standard deviations discrepancy remains.

If the 2-sigma effect persists or increases when more data is
collected and analysed by CDF, experimental effort to re-address the
estimation of the uncertainties might be acknowledged. Also more
theoretical work is needed to further improve predictions, e.g.~by
matching the NLO t-channel single top calculation in the 4-flavor
scheme to a parton shower program following e.g.~the
MC@NLO~\cite{Frixione:2002ik} or POWHEG~\cite{Nason:2004rx} methods.

\section{CONCLUSIONS}
There are three measurements in the top quark sector that are about 2
standard deviations away from its predictions within the Standard
Model:

\begin{itemize}
\item There is a small excess of events in the tail of the $H_T$
  distribution. Possible explanations might be that for the $H_T$
  observable the tail of the distribution is known to be non-trivial
  to model using Monte Carlo event generators. In particular, the
  uncertainty from higher order corrections can be large. However, if
  the effect is truly there, a possible explanation is a fourth
  generation of quarks.
\item The top quark charge asymmetry is found to be more pronounced
  than predicted by NLO QCD, which is the first order to give a
  non-zero result. The question arises if higher order effects might
  be important here, and it has been shown by using resummation
  techniques that this is probably not the case for the (soft) real
  emission. However, given that at NLO the virtual corrections give a
  larger contribution than the real emission, also the virtual
  corrections should be calculated at higher order to provide a
  prediction at high enough precision.
\item The discrimination of s- and t-channel events in single top
  production by CDF. Even though in the Monte Carlo predictions used
  in the CDF analyses there were half as many spectator-$b$ quarks as
  compared to the recent NLO calculation, it does not explain the
  difference with D\O~nor with the theoretical predictions. There is
  no understanding yet of this peculiarity.
\end{itemize}

On the other hand, these three 2-sigma deviations could very well be
statistical fluctuations. It is therefore important to see what this
will give with more data, and what the LHC will tell is in the near
future.

\bibliography{database,conf_notes}{}

\begin{thebibliography}{10}

\bibitem{Abe:1995hr}
{\bf CDF} Collaboration, F.~Abe {\em et~al.}, Phys. Rev. Lett. {\bf 74}, 2626
  (1995), [\href{http://arXiv.org/abs/arXiv/hep-ex/9503002}{hep-ex/9503002}].

\bibitem{Abachi:1995iq}
{\bf D\O} Collaboration, S.~Abachi {\em et~al.}, Phys. Rev. Lett. {\bf 74},
  2632 (1995),
  [\href{http://arXiv.org/abs/arXiv/hep-ex/9503003}{hep-ex/9503003}].

\bibitem{:1900yx}
{\bf CDF \& D\O} Collaboration,
  \href{http://arXiv.org/abs/1007.3178}{arXiv:1007.3178 [hep-ex]}.

\bibitem{note9913}
{\bf CDF} Collaboration, Conf. Note 9913  (2009).

\bibitem{note6038}
{\bf D\O} Collaboration, Conf. Note 6038  (2010).

\bibitem{Aaltonen:2009iz}
{\bf CDF} Collaboration, T.~Aaltonen {\em et~al.},
  \href{http://arXiv.org/abs/0903.2850}{arXiv:0903.2850 [hep-ex]}.

\bibitem{note5882}
{\bf D\O} Collaboration, Conf. Note 5882  (2009).

\bibitem{note10185}
{\bf CDF} Collaboration, Conf. Note 10185  (2010).

\bibitem{note6062}
{\bf D\O} Collaboration, Conf. Note 6062  (2010).

\bibitem{note10035}
{\bf CDF} Collaboration, Conf. Note 10035  (2010).

\bibitem{note6034}
{\bf D\O} Collaboration, Conf. Note 6034  (2010).

\bibitem{Acosta:2005hr}
{\bf CDF} Collaboration, D.~E. Acosta {\em et~al.}, Phys. Rev. Lett. {\bf 95},
  102002 (2005),
  [\href{http://arXiv.org/abs/arXiv/hep-ex/0505091}{hep-ex/0505091}].

\bibitem{Abazov:2008yn}
{\bf D\O} Collaboration, V.~M. Abazov {\em et~al.}, Phys. Rev. Lett. {\bf 100},
  192003 (2008), [\href{http://arXiv.org/abs/0801.1326}{arXiv:0801.1326
  [hep-ex]}].

\bibitem{note10211}
{\bf CDF} Collaboration, Conf. Note 10211  (2010).

\bibitem{note5722}
{\bf D\O} Collaboration, Conf. Note 5722  (2008).

\bibitem{note5950}
{\bf D\O} Collaboration, Conf. Note 5950  (2009).

\bibitem{:2007kq}
{\bf CDF} Collaboration, T.~Aaltonen {\em et~al.}, Phys.Rev. {\bf D78}, 111101
  (2008), [\href{http://arXiv.org/abs/0712.3273}{arXiv:0712.3273 [hep-ex]}].

\bibitem{note9939}
{\bf CDF} Collaboration, Conf. Note 9939  (2010).

\bibitem{Abazov:2006vd}
{\bf D\O} Collaboration, V.~Abazov {\em et~al.}, Phys.Rev.Lett. {\bf 98},
  041801 (2007),
  [\href{http://arXiv.org/abs/arXiv/hep-ex/0608044}{hep-ex/0608044}].

\bibitem{note10173}
{\bf CDF} Collaboration, Conf. Note 10173  (2010).

\bibitem{Abazov:2009xq}
{\bf D\O} Collaboration, V.~Abazov {\em et~al.}, Phys.Rev.Lett. {\bf 103},
  132001 (2009), [\href{http://arXiv.org/abs/0906.1172}{arXiv:0906.1172
  [hep-ex]}].

\bibitem{Group:2009qk}
{\bf CDF \& D\O} Collaboration, T.~E.~W. Group,
  \href{http://arXiv.org/abs/0908.2171}{arXiv:0908.2171 [hep-ex]}.

\bibitem{Aaltonen:2010jr}
{\bf CDF} Collaboration, T.~Aaltonen {\em et~al.},
  \href{http://arXiv.org/abs/1004.1181}{arXiv:1004.1181 [hep-ex]}.

\bibitem{Abazov:2009pa}
{\bf D\O} Collaboration, V.~M. Abazov {\em et~al.}, Phys.Lett. {\bf B682}, 363
  (2010), [\href{http://arXiv.org/abs/0907.4259}{arXiv:0907.4259 [hep-ex]}].

\bibitem{note10110}
{\bf CDF} Collaboration, Conf. Note 10110  (2010).

\bibitem{note5892}
{\bf D\O} Collaboration, Conf. Note 5892  (2010).

\bibitem{Abazov:2009ps}
{\bf D\O} Collaboration, V.~Abazov {\em et~al.}, Phys.Lett. {\bf B674}, 4
  (2009), [\href{http://arXiv.org/abs/0901.1063}{arXiv:0901.1063 [hep-ex]}].

\bibitem{note10104}
{\bf CDF} Collaboration, Conf. Note 10104  (2010).

\bibitem{:2009zh}
{\bf D\O} Collaboration, V.~Abazov {\em et~al.}, Phys.Lett. {\bf B682}, 278
  (2009), [\href{http://arXiv.org/abs/0908.1811}{arXiv:0908.1811 [hep-ex]}].

\bibitem{Aaltonen:2009qu}
{\bf CDF} Collaboration, T.~Aaltonen {\em et~al.}, Phys.Rev.Lett. {\bf 103},
  041801 (2009), [\href{http://arXiv.org/abs/0902.3276}{arXiv:0902.3276
  [hep-ex]}].

\bibitem{Abazov:2008vj}
{\bf D\O} Collaboration, V.~Abazov {\em et~al.}, Phys.Rev.Lett. {\bf 100},
  211803 (2008), [\href{http://arXiv.org/abs/0803.3256}{arXiv:0803.3256
  [hep-ex]}].

\bibitem{Abazov:2008rn}
{\bf D\O} Collaboration, V.~Abazov {\em et~al.}, Phys.Rev.Lett. {\bf 102},
  191802 (2009), [\href{http://arXiv.org/abs/0807.0859}{arXiv:0807.0859
  [hep-ex]}].

\bibitem{Abazov:2009ky}
{\bf D\O} Collaboration, V.~Abazov {\em et~al.}, Phys.Rev.Lett. {\bf 102},
  092002 (2009), [\href{http://arXiv.org/abs/0901.0151}{arXiv:0901.0151
  [hep-ex]}].

\bibitem{Aaltonen:2008qr}
{\bf CDF} Collaboration, T.~Aaltonen {\em et~al.}, Phys.Rev.Lett. {\bf 102},
  151801 (2009), [\href{http://arXiv.org/abs/0812.3400}{arXiv:0812.3400
  [hep-ex]}].

\bibitem{Abazov:2010qk}
{\bf D\O} Collaboration, V.~M. Abazov {\em et~al.},
  \href{http://arXiv.org/abs/1006.3575}{arXiv:1006.3575 [hep-ex]}.

\bibitem{:2008aaa}
{\bf CDF} Collaboration, T.~Aaltonen {\em et~al.}, Phys.Rev.Lett. {\bf 101},
  192002 (2008), [\href{http://arXiv.org/abs/0805.2109}{arXiv:0805.2109
  [hep-ex]}].

\bibitem{Rubin:2010xp}
M.~Rubin, G.~P. Salam and S.~Sapeta,
  \href{http://arXiv.org/abs/1006.2144}{arXiv:1006.2144 [hep-ph]}.

\bibitem{Alwall:2008qv}
J.~Alwall, S.~de~Visscher and F.~Maltoni, JHEP {\bf 0902}, 017 (2009),
  [\href{http://arXiv.org/abs/0810.5350}{arXiv:0810.5350 [hep-ph]}].

\bibitem{Catani:2001cc}
S.~Catani, F.~Krauss, R.~Kuhn and B.~Webber, JHEP {\bf 0111}, 063 (2001),
  [\href{http://arXiv.org/abs/arXiv/hep-ph/0109231}{hep-ph/0109231}].

\bibitem{Alwall:2007fs}
J.~Alwall {\em et~al.}, Eur.Phys.J. {\bf C53}, 473 (2008),
  [\href{http://arXiv.org/abs/0706.2569}{arXiv:0706.2569 [hep-ph]}].

\bibitem{Kuhn:1998jr}
J.~H. Kuhn and G.~Rodrigo, Phys.Rev.Lett. {\bf 81}, 49 (1998),
  [\href{http://arXiv.org/abs/arXiv/hep-ph/9802268}{hep-ph/9802268}].

\bibitem{Kuhn:1998kw}
J.~H. Kuhn and G.~Rodrigo, Phys.Rev. {\bf D59}, 054017 (1999),
  [\href{http://arXiv.org/abs/arXiv/hep-ph/9807420}{hep-ph/9807420}].

\bibitem{Almeida:2008ug}
L.~G. Almeida, G.~F. Sterman and W.~Vogelsang, Phys.Rev. {\bf D78}, 014008
  (2008), [\href{http://arXiv.org/abs/0805.1885}{arXiv:0805.1885 [hep-ph]}].

\bibitem{Ahrens:2010zv}
V.~Ahrens, A.~Ferroglia, M.~Neubert, B.~D. Pecjak and L.~L. Yang,
  \href{http://arXiv.org/abs/1003.5827}{arXiv:1003.5827 [hep-ph]}.

\bibitem{Cao:2010zb}
Q.-H. Cao, D.~McKeen, J.~L. Rosner, G.~Shaughnessy and C.~E.~M. Wagner, Phys.
  Rev. {\bf D81}, 114004 (2010),
  [\href{http://arXiv.org/abs/1003.3461}{arXiv:1003.3461 [hep-ph]}].

\bibitem{Bonciani:2008az}
R.~Bonciani, A.~Ferroglia, T.~Gehrmann, D.~Maitre and C.~Studerus, JHEP {\bf
  0807}, 129 (2008), [\href{http://arXiv.org/abs/0806.2301}{arXiv:0806.2301
  [hep-ph]}].

\bibitem{Bonciani:2009nb}
R.~Bonciani, A.~Ferroglia, T.~Gehrmann and C.~Studerus, JHEP {\bf 0908}, 067
  (2009), [\href{http://arXiv.org/abs/0906.3671}{arXiv:0906.3671 [hep-ph]}].

\bibitem{Korner:2008bn}
J.~G. Korner, Z.~Merebashvili and M.~Rogal, Phys. Rev. {\bf D77}, 094011
  (2008), [\href{http://arXiv.org/abs/0802.0106}{arXiv:0802.0106 [hep-ph]}].

\bibitem{Dittmaier:2007wz}
S.~Dittmaier, P.~Uwer and S.~Weinzierl, Phys. Rev. Lett. {\bf 98}, 262002
  (2007), [\href{http://arXiv.org/abs/arXiv/hep-ph/0703120}{hep-ph/0703120}].

\bibitem{Czakon:2008zk}
M.~Czakon, Phys.Lett. {\bf B664}, 307 (2008),
  [\href{http://arXiv.org/abs/0803.1400}{arXiv:0803.1400 [hep-ph]}].

\bibitem{GehrmannDeRidder:2005cm}
A.~Gehrmann-De~Ridder, T.~Gehrmann and E.~W.~N. Glover, JHEP {\bf 09}, 056
  (2005), [\href{http://arXiv.org/abs/arXiv/hep-ph/0505111}{hep-ph/0505111}].

\bibitem{Catani:2007vq}
S.~Catani and M.~Grazzini, Phys. Rev. Lett. {\bf 98}, 222002 (2007),
  [\href{http://arXiv.org/abs/arXiv/hep-ph/0703012}{hep-ph/0703012}].

\bibitem{Czakon:2010td}
M.~Czakon, Phys.Lett. {\bf B693}, 259 (2010),
  [\href{http://arXiv.org/abs/1005.0274}{arXiv:1005.0274 [hep-ph]}].

\bibitem{Somogyi:2005xz}
G.~Somogyi, Z.~Trocsanyi and V.~Del~Duca, JHEP {\bf 0506}, 024 (2005),
  [\href{http://arXiv.org/abs/arXiv/hep-ph/0502226}{hep-ph/0502226}].

\bibitem{Abazov:2009ii}
{\bf D\O} Collaboration, .~V.~M. Abazov,
  \href{http://arXiv.org/abs/0903.0850}{arXiv:0903.0850 [hep-ex]}.

\bibitem{Aaltonen:2009jj}
{\bf CDF} Collaboration, T.~Aaltonen {\em et~al.},
  \href{http://arXiv.org/abs/0903.0885}{arXiv:0903.0885 [hep-ex]}.

\bibitem{Boos:2006af}
E.~E. Boos, V.~E. Bunichev, L.~V. Dudko, V.~I. Savrin and A.~V. Sherstnev,
  Phys. Atom. Nucl. {\bf 69}, 1317 (2006).

\bibitem{Campbell:2009gj}
J.~M. Campbell, R.~Frederix, F.~Maltoni and F.~Tramontano, JHEP {\bf 10}, 042
  (2009), [\href{http://arXiv.org/abs/0907.3933}{arXiv:0907.3933 [hep-ph]}].

\bibitem{Campbell:2009ss}
J.~M. Campbell, R.~Frederix, F.~Maltoni and F.~Tramontano, Phys. Rev. Lett.
  {\bf 102}, 182003 (2009),
  [\href{http://arXiv.org/abs/0903.0005}{arXiv:0903.0005 [hep-ph]}].

\bibitem{Binoth:2010ra}
{\bf SM and NLO Multileg Working Group} Collaboration, J.~R. Andersen {\em
  et~al.}, \href{http://arXiv.org/abs/1003.1241}{arXiv:1003.1241 [hep-ph]}.

\bibitem{Sullivan:2004ie}
Z.~Sullivan, Phys. Rev. {\bf D70}, 114012 (2004),
  [\href{http://arXiv.org/abs/arXiv/hep-ph/0408049}{hep-ph/0408049}].

\bibitem{lueck}
{\bf CDF} Collaboration, J.~Lueck, Private Communication .

\bibitem{Frixione:2002ik}
S.~Frixione and B.~R. Webber, JHEP {\bf 06}, 029 (2002),
  [\href{http://arXiv.org/abs/arXiv/hep-ph/0204244}{hep-ph/0204244}].

\bibitem{Nason:2004rx}
P.~Nason, JHEP {\bf 11}, 040 (2004),
  [\href{http://arXiv.org/abs/arXiv/hep-ph/0409146}{hep-ph/0409146}].

\end{thebibliography}
\bibliographystyle{MyStyle}

\end{document}